\global\def\draftcontrol{0}

   \def\versionno{From Courant to actions}

\catcode`\@=11

\expandafter\ifx\csname draftcontrol\endcsname\relax\global\def\draftcontrol{0}
\fi

{\count255=\time\divide\count255 by 60
\xdef\hourmin{\number\count255}
\multiply\count255 by-60\advance\count255 by\time
\xdef\hourmin{\hourmin:\ifnum\count255<10 0\fi\the\count255}}
\def\draftdate{\number\month/\number\day/\number\year\ \ \ \hourmin }

\newcommand\makepapertitle{\par
  \begingroup
    \renewcommand\thefootnote{\@fnsymbol\c@footnote}%
    \def\@makefnmark{\rlap{\@textsuperscript{\normalfont\@thefnmark}}}%
    \long\def\@makefntext##1{\parindent 1em\noindent
            \hb@xt@1.8em{%
                \hss\@textsuperscript{\normalfont\@thefnmark}}##1}%
     \newpage
     \global\@topnum\z@   
     \@makepapertitle
     \thispagestyle{empty}\@thanks
  \endgroup
  \setcounter{footnote}{0}%
  \global\let\thanks\relax
  \global\let\makepapertitle\relax
  \global\let\@makepapertitle\relax
  \global\let\@thanks\@empty
  \global\let\@author\@empty
  \global\let\@date\@empty
  \global\let\@title\@empty
  \global\let\title\relax
  \global\let\author\relax
  \global\let\date\relax
  \global\let\and\relax
  \def\version{\let\version\@version\@gobble}
}
\def\@makepapertitle{%
  \newpage
   \ifnum\draftcontrol=1 {}
   \version\versionno
   \vskip 3em%
   \else
   \hfill\hbox to 3cm {\parbox{4cm}{\@pubnum}\hss}%
   \vskip 3em%
   \fi
   \begin{center}%
   \let \footnote \thanks
     {\LARGE {\@title}}%
     \vskip 1.5em%
     {\normalsize
       \lineskip .5em%
       \begin{tabular}[t]{c}%
         \@author
       \end{tabular}\par}%
     \vskip 1.5em%
     {\@bstract}%
     \end{center}%
     \vskip 1.5em
     \@date%
   \par
}

\gdef\@pubnum{}
\def\pubnum#1{%
  \gdef\@pubnum{#1}}

\gdef\@bstract{}
\def\Abstract#1{%
  \gdef\@bstract{%
   \parbox{\textwidth-0pc}{%
   \centerline{\bf Abstract}\penalty1000%
\noindent
\renewcommand\baselinestretch{1.0}%
{#1}}}
}

\def\ps@paper{\let\@mkboth\@gobbletwo%
     \ifnum\draftcontrol=1
        \def\@oddfoot{\hbox to \textwidth{\tiny \versionno \hfil\tiny\draftdate}%
        \hskip -\textwidth \hbox to \textwidth{\hfil\rm\thepage\hfil}}%
     \else\def\@oddfoot{\hbox to \textwidth{\hfil\rm\thepage\hfil}}
     \fi
     \let\@evenfoot\@oddfoot
}

\def\@version#1{\ifnum\draftcontrol=1
\typeout{}\typeout{#1}\typeout{}
\vskip3mm\centerline{\hbox{\fbox{\normalsize{\tt DRAFT -- #1 -- }
                   {\draftdate}}}}\vskip3mm
\fi}
\let\version\@version
\long\def\eqlabel#1{\ifnum\draftcontrol=1
                    \tag@false  
                    \tag*{(\theequation) \hbox to -0.2cm{\hspace{0cm}\small{#1}\hss}}
                    \refstepcounter{equation}
                    \edef\@currentlabel{\theequation}
                    \ltx@label{#1}          
                    \else
                    \label{#1}
                    \fi
                    }
\let\st@bibitem\@bibitem
\let\st@lbibitem\@lbibitem
\ifnum\draftcontrol=1
  \def\@bibitem#1{%
    \st@bibitem{#1}\a@@label{#1}\ignorespaces}
  \def\@lbibitem[#1]#2{%
    \st@lbibitem[#1]{#2}\a@@label{#2}\ignorespaces}
  \def\a@@label#1{%
    \gdef\a@lab{\smash{\normalfont\small#1}}
    \ifvmode
      \if@inlabel
        \global\setbox\@labels\hbox{%
          \llap{\a@lab\let\a@lab\relax
                \kern\@totalleftmargin\kern\marginparsep}%
          \box\@labels}%
      \fi
    \fi}
\fi

\documentclass[12pt,letterpaper]{article}

\usepackage{amsmath,amssymb,array,calc,rotating,epsfig,psfrag}
\usepackage[nosort]{cite}
\usepackage{graphicx}
\usepackage{hyperref}
\usepackage{mathtools,slashed}

\ifnum\draftcontrol=1
\fi

\renewcommand\baselinestretch{1.25}
\renewcommand\section{\@startsection {section}{1}{\z@}%
                                   {-3.5ex \@plus -1ex \@minus -.2ex}%
                                   {2.3ex \@plus.2ex}%
                                   {\normalfont\large\bfseries}}
\renewcommand\subsection{\@startsection{subsection}{2}{\z@}%
                                   {-3.25ex\@plus -1ex \@minus -.2ex}%
                                   {1.5ex \@plus .2ex}%
                                   {\normalfont\normalsize\bfseries}}
\renewcommand\subsubsection{\@startsection{subsubsection}{3}{\z@}%
                                   {-3.25ex\@plus -1ex \@minus -.2ex}%
                                   {1.5ex \@plus .2ex}%
                                   {\normalfont\normalsize\it}}
\renewcommand\paragraph{\@startsection{paragraph}{4}{\z@}%
                                   {-3.25ex\@plus -1ex \@minus -.2ex}%
                                   {1.5ex \@plus .2ex}%
                                   {\normalfont\normalsize\bf}}

\def\revise#1       {\raisebox{-0em}{\rule{3pt}{1em}}%
                     \marginpar{\raisebox{.5em}{\vrule width3pt\
                     \vrule width0pt height 0pt depth0.5em
                     \hbox to 0cm{\hspace{0cm}{%
                     \parbox[t]{4em}{\raggedright\footnotesize{#1}}}\hss}}}}

\def\del          {\partial}

\def\tr           {\mathop{\rm Tr}}

\def\half{{\frac12}}

\def\sqr#1#2{{\vcenter{\vbox{\hrule height.#2pt
 \hbox{\vrule width.#2pt height#1pt \kern#1pt
 \vrule width.#2pt}\hrule height.#2pt}}}}

\def\a{\alpha}
\def\b{\beta}
\def\r{\rho}

\def\O{\Omega}

\def\o{\omega}

\def\m{\mu}
\def\g{\gamma}
\def\l{\lambda}
\def\n{\nu}
\def\bn{\bar{\nu}}
\def\bm{\bar{\mu}}

\def\stackunder#1#2{\mathrel{\mathop{#2}\limits_{#1}}}%
\catcode`\@=12

\usepackage[active]{srcltx}
\usepackage{color}

\begin{document}




\newcommand{\be}{\begin{equation}}
\newcommand{\ee}{\end{equation}}
\newcommand{\beq}{\begin{equation}}
\newcommand{\eeq}{\end{equation}}
\newcommand{\ba}{\begin{eqnarray}}
\newcommand{\ea}{\end{eqnarray}}
\newcommand{\nn}{\nonumber}

\def\vol{\bf vol}
\def\Vol{\bf Vol}
\def\del{{\partial}}
\def\vev#1{\left\langle #1 \right\rangle}
\def\cn{{\cal N}}
\def\co{{\cal O}}
\def\IC{{\mathbb C}}
\def\IR{{\mathbb R}}
\def\IZ{{\mathbb Z}}
\def\RP{{\bf RP}}
\def\CP{{\bf CP}}
\def\Poincare{{Poincar\'e }}
\def\tr{{\rm tr}}
\def\tp{{\tilde \Phi}}
\def\Y{{\bf Y}}
\def\te{\theta}
\def\bX{\bf{X}}

\def\TL{\hfil$\displaystyle{##}$}
\def\TR{$\displaystyle{{}##}$\hfil}
\def\TC{\hfil$\displaystyle{##}$\hfil}
\def\TT{\hbox{##}}
\def\HLINE{\noalign{\vskip1\jot}\hline\noalign{\vskip1\jot}} 
\def\seqalign#1#2{\vcenter{\openup1\jot
  \halign{\strut #1\cr #2 \cr}}}
\def\lbldef#1#2{\expandafter\gdef\csname #1\endcsname {#2}}
\def\eqn#1#2{\lbldef{#1}{(\ref{#1})}%
\begin{equation} #2 \label{#1} \end{equation}}
\def\eqalign#1{\vcenter{\openup1\jot   }}
\def\eno#1{(\ref{#1})}
\def\href#1#2{#2}
\def\half{{1 \over 2}}

\def\ads{{\it AdS}}
\def\adsp{{\it AdS}$_{p+2}$}
\def\cft{{\it CFT}}

\newcommand{\ber}{\begin{eqnarray}}
\newcommand{\eer}{\end{eqnarray}}

\newcommand{\bea}{\begin{eqnarray}}
\newcommand{\eea}{\end{eqnarray}}

\newcommand{\beqar}{\begin{eqnarray}}
\newcommand{\cN}{{\cal N}}
\newcommand{\cO}{{\cal O}}
\newcommand{\cA}{{\cal A}}
\newcommand{\cT}{{\cal T}}
\newcommand{\cF}{{\cal F}}
\newcommand{\cC}{{\cal C}}
\newcommand{\cR}{{\cal R}}
\newcommand{\cW}{{\cal W}}
\newcommand{\eeqar}{\end{eqnarray}}
\newcommand{\lm}{\lambda}\newcommand{\Lm}{\Lambda}
\newcommand{\eps}{\epsilon}


\newcommand{\nonu}{\nonumber}
\newcommand{\oh}{\displaystyle{\frac{1}{2}}}
\newcommand{\dsl}
  {\kern.06em\hbox{\raise.15ex\hbox{$/$}\kern-.56em\hbox{$\partial$}}}
\newcommand{\as}{\not\!\! A}
\newcommand{\ps}{\not\! p}
\newcommand{\ks}{\not\! k}
\newcommand{\D}{{\cal{D}}}
\newcommand{\dv}{d^2x}
\newcommand{\Z}{{\cal Z}}
\newcommand{\N}{{\cal N}}
\newcommand{\Dsl}{\not\!\! D}
\newcommand{\Bsl}{\not\!\! B}
\newcommand{\Psl}{\not\!\! P}
\newcommand{\eeqarr}{\end{eqnarray}}
\newcommand{\ZZ}{{\rm \kern 0.275em Z \kern -0.92em Z}\;}

\def\s{\sigma}
\def\a{\alpha}
\def\b{\beta}
\def\r{\backslash l}
\def\d{\delta}
\def\g{\gamma}
\def\G{\Gamma}
\def\ep{\epsilon}
\makeatletter \@addtoreset{equation}{section} \makeatother
\renewcommand{\theequation}{\thesection.\arabic{equation}}

\def\be{\begin{equation}}
\def\ee{\end{equation}}
\def\bea{\begin{eqnarray}}
\def\eea{\end{eqnarray}}
\def\m{\mu}
\def\n{\nu}
\def\g{\gamma}
\def\p{\phi}
\def\L{\Lambda}
\def \W{{\cal W}}
\def\bn{\bar{\nu}}
\def\bm{\bar{\mu}}
\def\bw{\bar{w}}
\def\ba{\bar{\alpha}}
\def\bb{\bar{\beta}}

\begin{titlepage}

\vskip 1.7 cm

\centerline{\bf \Large  Dynamical Projective  Curvature in  Gravitation }

\vskip 1.5cm 
\centerline{\bf \footnotesize Samuel Brensinger$^{a,}$\footnote{samuel-brensinger@uiowa.edu}and  Vincent G. J. Rodgers$^b,$\footnote{vincent-rodgers@uiowa.edu} }

\vskip 1cm
\centerline{${}^a$ Department of Mathematics}
\centerline{and}
\centerline{${}^b$ Department of Physics and Astronomy}
\centerline{ \it  The University of Iowa}
\centerline{\it Iowa City, IA 52242}

\vspace{1cm}

\begin{abstract}
By using a projective connection over the space of two-dimensional affine connections,  we are able to show that the metric  interaction of Polyakov 2D gravity with a coadjoint element arises naturally through the projective Ricci tensor.   Through the curvature invariants of Thomas and Whitehead, we are able to define an action that could describe dynamics to the projective connection.  We discuss implications of  the projective connection in higher dimensions  as related to gravitation.  
\end{abstract}
\end{titlepage}
\section{Summary of Paper}
In one dimension, the smooth circle and line are characterized by the algebra of vector fields, called the Virasoro algebra. It can be further endowed with more structure by adding a gauge group so that points on the circle are mapped into the group.   The Kac-Moody algebra locally characterizes this added structure. The symmetries of the circle lead to two dimensional physics through symplectic structures associated with the coadjoint orbits of the Virasoro algebra and Kac-Moody algebras.  The coadjoint elements (\(\mathcal{D}, A\)), corresponding to a quadratic differential \(\mathcal{D}\) and a gauge field A, are realized as a  cosmological term in the effective action of two-dimensional gravity  and background gauge field coupled to a WZW\ model, respectively.  The gauge field has meaning in any dimension and can be made dynamical by adding a Yang-Mills action in 2D and higher.  In this note we address two questions: 1)\ Is there a principle that can give  dimensional ubiquity for the field $\mathcal{D}$? and 2)  What would be the corresponding dynamical action?  In other words, by ``lowering" the  dimension to 2D,   are new gravitational degrees of freedom that are independent of Einstein now manifest and can they have an interpretation in higher dimensions? We answer these  questions by first recognizing that  \(\mathcal{D}\) is related to projective geometry, a notion that is well-defined in higher dimensions. Then by using the Thomas-Whitehead projective connection \( \tilde \nabla\),  we are able to describe a projective curvature invariant and build a dynamical  action for \( \mathcal{D}\) modelled on the Gauss-Bonnet action in Riemannian geometry.  We discuss the gravitational consequences of this action in two and four dimensions.            

\section{The Coadjoint Representation of the Virasoro Algebra}\label{intro}
\subsection*{The Virasoro Algebra and Its Dual}
 In this section we present a self-contained  review of the Virasoro algebra, its coadjoint representation, its coadjoint orbits and the relation of coadjoint elements to projective structure on the circle (or line).  The Virasoro algebra  is the unique central extension of the Witt algebra, which is the algebra of vectors on the circle $S^1$. Let $\xi$  be a vector on a circle parameterized by a coordinate \( \theta\). This can be written as 
\begin{equation} \xi^a(\theta)\partial_a= \xi(\theta) \frac{d}{d \theta}.\end{equation} 
Then the Witt algebra between two vector fields \( \xi\) and \(\eta\) can be defined through the Lie derivative,  
\begin{equation}
{\cal L}_{\xi} \eta^a = -\xi^b \partial_b \eta^a + \eta^b
\partial_b\xi^a = (\xi \circ \eta)^a,
\end{equation}
so that the  algebra of vector fields is defined through,
\begin{equation}
 [{\xi}, {\eta}] = {\xi \circ \eta}. 
\end{equation}
We can centrally extend this
algebra by denoting \( i\) as the central element and  writing a two-tuple, \( (\xi,a)\) to denote
\begin{equation}( \xi, a) = \xi(\theta) \frac{d}{d \theta} - ia, 
\end{equation}  where \( a \) is a real number. The centrally extended algebra for elements \( (\xi, a)\) and \( (\eta, b) \) then becomes, 
\begin{equation}
 [{(\xi}, a), ({\eta},b) ] = ({\xi \circ \eta}, 0), \label{algebra1}
\end{equation}
where $a$ and $b$ do not contribute and no new central component appears. However, one can introduce a two-cocycle, \( ((*,*)) \), which maps two vector fields, say  \(\xi\) and \(\eta\) into a real number, $((\xi,\eta))_{0}$,  
\begin{equation}
((\xi,\eta))_0 = \frac{c}{2\pi} \int (\xi \eta''' )\,d\theta,
\label{Gelfand-Fuchs}
\end{equation}
where $c$ is a constant and the $'$ denotes a \( \theta \) derivative.
This two-cocycle  is anti-symmetric in \( \xi \leftrightarrow\ \eta\) and satisfies the Jacobi identity since
\begin{equation} 
((\xi,[\eta,\alpha]))_{0}+ ((\alpha,[\xi,\eta]))_{0}+ ((\eta,[\alpha,\xi]))_{0} =0.
\end{equation}
Using this, one has a  centrally extended algebra,  \   \begin{equation}
 [({\xi}, a), ({\eta},b) ] = ({\xi \circ \eta}, ((\xi,\eta))_{0}), \label{adjoint}
\end{equation}
that is still a Lie algebra, called the Virasoro algebra.   The two-cocycle, $((\xi,\eta))_{0}$, is the  Gelfand-Fuchs cocycle. One may see it as a pairing of the vector $\xi$ with a one-cocycle of $\eta$, where this one-cocycle arises from a  projective transformation that has mapped the    vector field $\eta$  into a quadratic differential,\begin{equation}
\eta \partial_\theta  \rightarrow \eta''' d\theta^2. \label{mapping}
\end{equation}  Through this pairing, one has formed the \emph{dual} of the algebra, \(\mathcal{G}^*\), where the elements are quadratic differentials. One goes further by introducing the coadjoint representation on the dual space of the algebra.  Coadjoint elements of the Virasoro algebra  are represented by $(B, c),$ which is  the direct sum of the dual of  the algebra (the element \(B\)) and the reals  (the element $c$).   An invariant pairing   between \((\xi, a)\) and \((B,c)\), $<(\xi,a) | (B, c)>$,  is then defined by
\begin{equation}
< (\xi,a) | (B, c) > \equiv \int (\xi B) d \theta + a c.
\end{equation}
  Using the invariance, we act on  the pairing with an element of the   algebra, say $(\eta, d)$, 
\begin{equation}
(\eta, d)\ast< (\xi,a) | (B, c) > =0,\end{equation}
to find the coadjoint action of the algebra via the Leibnitz rule.  Indeed
\begin{equation} < ad_{(\eta,d)}(\xi,a) | (B,c) >+< (\xi,a) |ad^*_{(\eta, d)}( B, c)> =0, \label{Leibnitz}
\end{equation}
implies that the coadjoint action of \((\eta, d) \) on \((B, c)  \) is ,  
\be
ad^*_{(\eta,d)}( B, c)=(\eta B' + 2 \eta' B -  c\,\ \eta''',0). \label{coadjointaction}
\ee
In Eq.(\ref{Leibnitz}), \(ad_{(\eta, d)}(\xi,a)\) is the adjoint action of centrally extended vector fields on themselves and is given by Eq.(\ref{adjoint}). \subsection*{Relation to Sturm-Liouville and Projective Structures}
The first two summands of Eq.(\ref{coadjointaction}) in the left component of the two-tuple correspond to the  Lie derivative of a quadratic differential. However, the last summand is the Gelfand-Fuchs cocycle, making the coadjoint action an affine module \cite{Tabachnikov,OvsienkoValentin2005Pdgo}.  Kirillov \cite{Kirillov:1982kav} observed that this action is the same as the action of vector fields on the space of Sturm-Liouville operators so that we have the correspondence,  
\be 
(B, c) \Leftrightarrow -2 c \frac{d^2}{dx^2} + B(x),
\label{correspondence}\ee
where on the left side $(B, c)$ is identified with a centrally extended coadjoint element of the Virasoro algebra and on the right side  is a Sturm-Liouville operator and $B(x)$ is a Strum-Liouville potential. 

To see this correspondence \cite{Tabachnikov,OvsienkoValentin2005Pdgo}, let
 \(\phi_A\) and  \(\phi_B\) be two independent solutions  of the Sturm-Liouville equation,  
\begin{equation}
 (-2 c \frac{d^2}{dx^2} + B(x)) \phi_A=0, \hskip.3in
 (-2 c \frac{d^2}{dx^2} + B(x)) \phi_B=0. \label{SturmLiouville2}
\end{equation}
 The solution space is two dimensional so it can be  spanned by \(\phi_A\) and  \(\phi_B\).  Define a function \(f(x)= \frac{\phi_A(x)}{\phi_B(x)}\).  Then 
\begin{equation}
f'(x) = \frac{\phi_A'(x)\phi_B(x)-\phi_A(x)\phi'_B(x)}{\phi_B(x)^2},
\end{equation} where we recognize the numerator as  the Wronskian 
of the two independent solutions. This guarantees that   \(f'(x) \neq 0\). Similarly, \begin{equation}
f''(x) = -2 \frac{\phi_A'(x) \phi_B'(x)}{\phi_B(x)^2}+ 2 \frac{\phi_A(x) \phi_B'(x)^2}{\phi_B(x)^3}+ \frac{\phi_A''(x)}{\phi_B(x)}- \frac{\phi_A(x) \phi_B''(x)}{\phi_B(x)^2} \label{secondDerivative}
\end{equation} 
which becomes 
\begin{equation}
f''(x) =-2 \frac{\phi_A'(x) \phi_B'(x)}{\phi_B(x)^2}+ 2 \frac{\phi_A(x) \phi_B'(x)^2}{\phi_B(x)^3}
\end{equation} after using Eq.(\ref{SturmLiouville2}). 
Continuing in this manner,  
\begin{equation}
f'''(x) = \frac{2\left(\phi_B(x) \phi_A'(x)- \phi_A(x) \phi_B'(x)\right)(B(x) \phi_B(x)^2 + 3   c\,\  \phi_B'(x)^2)}{ c \,\phi_B(x)^4 }.
\end{equation}
One can then extract \(B(x)\) as a particular ratio of derivatives,
\begin{equation}
B(x) = \frac{ c}{2} \left( \frac{f'''(x)}{f^{'}(x)}- \frac{3}{2} \left(\frac{f''(x)}{f^{'}(x)}\right)^2\right)=S(f(x).
\end{equation} This is precisely the Schwarzian derivative of \(f(x) \) with respect to \(x\), \(S(f(x)).  \) 
  We may therefore consider \(f(x) \) to be an affine parameter \(\tau \equiv f(x)\) on the projective line \(\mathbb{P}^1\). 
We then have a one parameter family of Strum-Liouville operators,
 \begin{displaymath}
L_\tau \phi  = -2 c \frac{d^2}{dx^2} \phi + B_\tau \phi =0. \label{tauSturm}
\end{displaymath}
   This is invariant under the action of the vector field  $(\eta, d) $, so we have 
\begin{displaymath}
(\eta,d)*L_\tau \phi  = (ad^*_{(\eta,d)}\,L_\tau )\phi +L_\tau ( (\eta,d)*\phi)=0, 
\end{displaymath}   where the coadjoint action, Eq,(\ref{coadjointaction}) defines the transformation of the operator \(L_\tau \), and where  the solution \(\phi\) transforms as a  scalar density of weight \(-\frac{1}{2}\), i.e. \[\mathcal{L}_{(\eta,d)} \phi = \eta \phi' - \frac{1}{2} \eta' \phi.\] Again the central element $d$ does nothing.  In this way we are able to make direct contact with one-dimensional projective geometry.  In other words the invariants of the Virasoro algebra can be identified with invariants of Sturm-Liouville, which in turn defines a projective structure on $S^1$.  This verifies Eq.(\ref{correspondence}), identifying ${ B}$ as a \emph{projective connection}\cite{Kirillov:1982kav}.   At the group level,  $B(x)$ transforms as a quadratic differential plus a Schwarzian derivative. The Schwarzian derivative is perhaps the most well-known and ubiquitous example of a projective differential geometry invariant\cite{Tabachnikov,OvsienkoValentin2005Pdgo}. In what follows, we will return to this correspondence by relating the invariant two-cocycle,
\begin{equation}
(\xi,\eta)_{(B, c)} = \frac{ c}{2\pi} \int (\xi \eta''' - \xi'''
  \eta)\,dx
+ \frac{1}{2\pi} \int (\xi \eta' - \xi' \eta )B\,dx,
\label{2cocycle2}
\end{equation}
 explicitly to the Thomas-Whitehead connection in Section \ref{TW2cocycle} below.   
\subsection*{Coadjoint Orbits}
One of the ways coadjoint orbits are  of interest to physicists  is with respect to string theory and gravity  through the quantization of the Virasoro algebra and the  anomalous contributions to the 2D effective quantum gravitational action. By studying  coadjoint orbits 
\cite{Witten:1987ty,Rai:1989js,Delius:1990pt,Bakas:1988mq,Alekseev:1988ce,Alekseev:1988vx}  ones gets both a geometric understanding of the underlying vacuum structure of gravity in 2D and the  anomalous contributions to gravitation.  By explicitly studying the field theory associated with the semi-direct product of  the Virasoro algebra and an affine Lie algebra  that defines a gauge theory,  one shows that the  
WZW \cite{Witten:1983tw} model
and Polyakov \cite{Polyakov:1987zb} 2D quantum gravity, derived via path integral quantization of  two-dimensional chiral fermions, are
 the geometric actions
associated with the affine Kac-Moody algebra and the Virasoro algebra
respectively.  

Each coadjoint orbit admits a natural symplectic two-form, $\O_{(B, c)}[*, *] $ on the Virasoro algebra, \cite{Kirillov:1982kav,Segal:1981ap}. Indeed using Eq.(\ref{2cocycle2}), we may define
\begin{equation}
\O_{(B, c)}[(\xi,a), (\eta,d)] =(\xi,\eta)_{(B, c)}.
\end{equation} 
The two-form is closed, via the Jacobi identity, and non-degenerate making it a natural symplectic two-form on the orbits defined by \((B, c)\). The isotropy algebra of \( (B,c) \), is defined by those vector fields,\( (\hat \eta , \hat d)\)  where
\begin{equation}
ad^*_{(\hat \eta, \hat d)}( B, c)=(\hat \eta B' + 2 {\hat \eta}' B -  c\,\ {\hat \eta}''',0)=(0,0).
\end{equation}
The subalgebra of all such vector fields generates the isotropy group, \(\mathrm{H}\), so that the coadjoint orbits are defined by \(\text{diff}S^{1}/\mathrm{H }\) where \(\text{diff}S^{1}\) denotes the Virasoro group. Coadjoint elements that can be related to each other through group action are in the same equivalence class and therefore live on the same orbit.  Because of this, the dual of the Virasoro algebra can be  foliated into distinct classical and quantum mechanical systems via the equivalence classes of the coadjoint elements, i.e. $[(B, c)]$.  This gives rise to a Hilbert space structure which is a direct sum of the equivalence classes\cite{Witten:1987ty}
\begin{equation}
\mathcal{H}=\oplus_i \mathcal{H}_{[B_i]}. 
\end{equation}  From a quantum mechanical viewpoint, the coadjoint elements determine the vacuum structure of their respective Hilbert spaces. 
By using this natural symplectic two form on coadjoint orbits  and a strategy for integrating the two-form \cite{Balachandran:1986hv,Balachandran:1979pc,Balachandran:1987st}, one also relates each orbit  specified by  coadjoint element  $(\mathcal{D},\tilde c)$ with a 2D gravitational action\cite{Rai:1989js}

\begin{equation}
S = \frac{\tilde c}{2 \pi} \int d x d \tau \left[
\frac{\partial^2_{x} s}{(\partial_{x}s)^2} \partial
_{\tau} \partial_{x} s - \frac{(\partial^2_{x}s)^2
(\partial_{\tau} s)}{(\partial_{x} s)^3} \right]
 - \int d x d \tau  \,\,\mathcal{D}(x)
\frac{(\partial s/ \partial \tau)}{(\partial s/ \partial
x)}. \label{Polyakov}
\end{equation}
Here $s(\theta; \l, \tau)$ corresponds to a two-parameter family of  elements of the Virasoro group.  Changing the  notation $\,x\rightarrow x_-$,$\tau \rightarrow x_+$, $s \rightarrow f,\,$ and $\mathcal{D}\rightarrow 0$, this action is identical to Polyakov's action\cite{Polyakov:1987zb} , viz
\begin{equation}
S= \frac{\tilde c}{2 \pi} \int d^2x  \left[ \left( \partial^2_- f\right) \left(\partial_+ \partial_- f\right)
\left( \partial_- f \right)^{-2} - \left(\partial^2_- f\right)^2
(\partial_+ f) \left( \partial_-
f\right)^{-3} \right] 
\end{equation}

where the gauge fixed metric is
\be
g_{a b} = \begin{pmatrix}0 & \frac{1}{2} \\
\frac{1}{2} & h_{++}(\theta,\tau) \\
\end{pmatrix},\ee and where $h_{++} = (\partial_+f/ \partial_-f)$  for a function $f(x_{-},x_+)$ in light-cone coordinates. 
The interaction term with the  coadjoint element $(\mathcal{D},0) $, viz. the last summand in Eq.(\ref{Polyakov}), admits a term
\be S_{(D,g)} = \int d^2x \,\mathcal{D}\,\, (\partial_+f/ \partial_-f) ,\label{interact}
\ee  
suggesting that $\mathcal{D}$ is the $\mathcal{D}=\mathcal{D}_{--}$ component of an external background field $\mathcal{D}_{\mu \nu}$ in two dimensions. Here it serves as a background cosmological term that influences the energy-momentum tensor.
\subsection*{The Quadratic Differential, \( \mathcal D_{a b} \), and Yang-Mills Potentials, \( A_a \)}
In Section \ref{TWPolyakov} we will show that by using a projective connection over the affine connections in two dimensions, Eq.(\ref{interact}) appears naturally as a background field in the Einstein-Hilbert in 2D.  We will go further and ascribe dynamics to this field in Section \ref{TWdynamics}.  This is akin to adding the Yang-Mills action to a WZW model that is coupled to a background gauge field $A_\m$.  Understanding the dynamics of the coadjoint elements in the Virasoro algebra will help to understand the stability of the quantization of orbits such as $\text{diff}S^{1}/SL^{(n)}(2,R)  $   as well as non-trivial cosmological contributions to gravitation in higher dimensions.  The partnership with the vector potential $A_\m$ (we will take $SU(M)$ as our gauge group) and the quadratic differential $\mathcal{D}_{\m \n}$ in describing the semi-direct product of the Kac-Moody (affine Lie algebra)  and the Virasoro algebra is another impetus for making $\mathcal{D}(\theta)$ dynamical.  

To explicitly show this partnership,   let us  review the relevant literature \cite{Lano:1994gx,Gates:2002xh}. As discussed above,  the coadjoint representation appears as the dual of  the adjoint 
representation through a  pairing between the two representations,  $<(\xi,a) | (B,c)>$.  We will consider the analogous pairing for the  algebra of the semi-direct product of the Virasoro algebra with an affine Lie algebra and its dual.  We write the mode decomposition of the Virasoro algebra  and the affine Lie algebra (Kac-Moody algebra) with structure constants \(f^{\a \b \gamma}\) on the circle by
\begin{equation}
\left[ L_N,L_M\right]  =(N-M)\,L_{N+M} 
+  c N^3\,\delta _{N+M,0} 
\end{equation}
for the Virasoro sector, 
\begin{equation}
\left[ J_N^\alpha ,J_M^\beta \right]  =i\,f^{\alpha \beta \gamma
}\,J_{N+M}^\beta +N\,k\;\delta _{N+M,0}\;\delta ^{\alpha \beta }
\end{equation}
for the affine Kac-Moody algebra, and 
\begin{equation}
\left[ L_N,J_M^\alpha \right]  = -M\;J_{N+M}^\alpha. 
\end{equation}
for the interacting commutation relations. Here $A, B, C \cdots Z$ are integers. To be explicit, the algebra may be realized on the circle by 
\begin{equation}
L_N= e_{N}^a \partial_a = ie^{iN\theta }\partial _\theta, \,\,\,\,\,\,\,J_N^\alpha 
= \tau ^\alpha e^{iN\theta },
\end{equation}so that a basis for the centrally extended algebra  can be written as
\begin{equation}
\left(L_A,J_B^\beta ,\rho\right).
\end{equation}
Here the last component, $\rho$,  is in the center of the algebra.  The adjoint representation acts  on itself through the commutation relations and explicitly is given by: 
\begin{equation}
\left( {L_A,J_B^\beta ,\rho }\right) {*}\left( {L_{N},J_{M}^{\alpha},\mu }\right)  =\left( {L_{new},J_{new},\b }%
\right) \label{case1-a}
\end{equation}
where
\begin{eqnarray}
L_{new}&=&  \,(A-N)\,L_{A+N} \cr
J_{new}&=&  -M J_{A+M}^{\alpha }+BJ_{B+N}^{\,\beta }+if\,^{\beta \alpha \lambda
}J_{B+M}^\lambda\cr
\beta \ &=&(c A^3) \delta _{A+N,0}+Bk\delta ^{\alpha
\beta }\delta _{B+M,0}. \label{case1-b}
\end{eqnarray}

In a similar way, we can construct a basis for the coadjoint representation. The coadjoint elements can be realized as $\left( \widetilde{L_N},\widetilde{J_M^\alpha },\widetilde{\mu }\right)$;

 \begin{equation}
{\tilde L}_N= e^{N}_{a b} dx^a dx^b = -ie^{-iN\theta }d\theta^2, \,\,\,\,\,\,\,{\tilde J}_N^\alpha 
= A_a^{N,\a} dx^a= \tau ^\alpha e^{-iN\theta },
\end{equation} and $\tilde \m$ is a constant. The $ e^{N}_{a b}$ are the components of a quadratic differential and the $A_a^{N,\a} $ are one-form components.  The pairing between the modes of the coadjoint representation and the adjoint representation is explicitly:

\begin{multline}
\left\langle   \left( L_A,J_B^\beta ,\rho \right)\left|\left( \widetilde{L_N},\widetilde{J_M^\alpha },\widetilde{\mu }%
\right)  \right.\right\rangle
= \frac{1}{2 \pi i}\int\left( e_{A}^ae^{N}_{a b} +\tr(J_B^\beta\ A_b^{M,\a})\right)   dx^b   +\rho \widetilde{\mu }\\= \frac{1}{2 \pi i}\int\left( e^{iA\theta } e^{-iN\theta }+\tr(\tau ^\beta e^{i B\theta }\tau ^\alpha e^{-iM\theta })\right)   d\theta   +\rho \widetilde{\mu }\\= \delta _{N,A}+\delta ^{\alpha \beta }\delta _{M,B}+\rho \widetilde{\mu },
\end{multline} 
where \(\tr(\tau^\alpha \tau^\beta)= \delta^{\alpha \beta}. \) Invariance of the pairing with respect to the action of the adjoint representation gives the transformation laws for the coadjoint representation  \cite{Lano:1994gx},
\begin{eqnarray}
\left( {L_A,J_B^\beta ,\rho }\right) &{*}&\left( {{\tilde L_N},{\tilde
J_M^\alpha },{\tilde \mu }}\right) =\left( {{\tilde L_{new}},{\tilde
J_M^\alpha },0}\right) \mbox{ } \,\,\,\,\,\,\,\mbox{\rm with,}\\ \cr
{{\tilde L_{new}}}{}{=}&& {(2A-N){\tilde L_{N-A}} {-} B\delta ^{\alpha \beta }{%
\tilde L_{M-B}}} -{\tilde \mu}(c A^3){\tilde
L_{-A}}\,\,\,\,\,\,\,\,\, \mbox{\rm and} \\\cr
{\tilde J_M^\alpha }{}{=} &&{(M-A){\tilde J_{M-A}^\alpha }-if\,^{\beta \nu
\alpha }{\tilde J_{M-B}^\nu }} -{\tilde \mu} B\,k\,{\tilde J_{-B}^{\,\beta }.%
} 
\label{case1-c}
\end{eqnarray}
In terms of the mode decomposition, the one-dimensional vector fields $\xi^j$ and the matrix valued gauge parameter $\Lambda^I_J$ are given by
\begin{equation}\xi^j = \sum_{N=-\infty}^\infty e^j_N \xi^N  \text{\,\,\,\,\,\,and\,\,\,\,\,\,\,\,\,} \L^I_J = \sum_{\a=1}^q \sum_{N=-\infty}^{\infty} \L^N_\a (J_N^\a)^I_J,
\end{equation}
where $q$ is the dimension of the affine Lie  algebra.  Then a generic member of the algebra may be written as the  three-tuple 
\begin{equation}{\cal F}=\left( \xi \left( \theta \right), \Lambda \left(
\theta \right) ,a\right)\label{eqn1}
\end{equation}
 containing a  one-dimensional vector field $\xi^i$,  an $M\times M$ matrix valued  gauge parameter $\L^I_J$, and a central
element $a$.  The pairing in the Virasoro sector is a  contraction of  a vector field \(\xi^i \) and a quadratic differential
 \(\mathcal{D}_{i j}\), 
 \begin{equation} <\xi,\mathcal{D}> = \int \xi^i \mathcal{D}_{i j} dx^j,
\end{equation}
 and the pairing in the Kac-Moody sector  of a gauge parameter $\L^I_J$ and a dual element $(A_i)^I_J$ is given by
 \be <\L,A> = \int \tr( \L A_{i}) dx^j.\ee
Therefore we may write
\begin{equation}
\mathcal{D} = \sum_{n=-\infty}^\infty \mathcal{D}^n {\tilde L}_n  \text{\,\,\,\,\,\,and\,\,\,\,\,\,\,\,\,} A = \sum_{\a=1}^M \sum_{n=-\infty}^{\infty} \L^n_\a \tilde{J}_n^\a.
\end{equation} 
 The coadjoint element is the three-tuple
\begin{equation} {\rm B} =\left( { \mathcal{D}}\left( \theta \right), {\rm A}\left(
\theta\right), \tilde \mu \right),\label{eqn2} 
\end{equation}
 which consists of a rank two
projective connection $\mathcal{D}_{a b}$, a gauge connection $A_a$ and a corresponding
central element $\mu$.  In this way the coadjoint action of an adjoint element $\mathcal{F}=\left( \xi \left( \theta \right), \Lambda \left(
\theta \right) ,a\right)$  
on a coadjoint element  $\mathcal{B}=\left( { \mathcal{D}}\left( \theta \right), { A}\left(
\theta \right),\tilde \mu \right) $ gives the transformation law \cite{Lano:1994gx}  
\begin{eqnarray}
\delta \widetilde{{\mathcal{B}}_F}&&=\left( \xi \left( \theta \right), \Lambda \left(
\theta \right) ,a\right) *\left( { \mathcal{D}}\left( \theta \right), { A}\left(
\theta \right),\tilde \mu \right) \cr
&& =\left( \delta { \mathcal{D}}\left( \theta \right), \delta { A}\left( \theta \right), 0\right).  
\end{eqnarray}
    The transformation laws now have the interpretation of a one-dimensional projective transformation on the projective connection \({ \mathcal{D}}\left( \theta \right)\) and gauge connection $A$, and the accompanying gauge transformations on these fields, i.e   
\begin{equation}
\delta { \mathcal{D}}\left( \theta \right) =\;
\stackunder{\rm coordinate\ transformation}{\underbrace{2\xi ^{^{\prime
}}{ \mathcal{D}}+{ \mathcal{D}}^{^{\prime }}\xi + \frac { c \tilde \mu}{2\pi}\xi^{\prime \prime
\prime } }}-\stackunder{\rm gauge\ transformation}
{\underbrace{\text{Tr}\left( { A}\Lambda^{\prime }\right)}} \label{variation_D}
\end{equation}
and
\begin{equation}
\delta { A(\theta )}=\;
\stackunder{\rm coordinate\ transformation}{\underbrace{{ A}^{\prime }\xi +\xi^{\prime }
{ A}}}-\stackunder{\rm gauge\ transformation}{\,\underbrace{[\Lambda
\,{ A}-{ A\,}\Lambda ]+k\,\tilde \mu \,\Lambda^{\prime
}}}. \label{variation_A}
\end{equation}
We note that \({\rm \mathcal{D}}\left( \theta \right)\) transforms inhomogeneously under the projective transformation, while  $A(\theta)$ transforms inhomogeneously under the gauge transformation. In terms of two-dimensional Yang-Mills, $A(\theta)$ may be regarded as the space component of a vector potential, $A_\mu =(A_\tau, A_\theta)$,  where one uses the temporal gauge to fix $A_\tau = 0$. When   Eq.(\ref{variation_A}) is set to zero and the  coordinate transformations are ignored, one sees that this parallels a Gauss Law constraint from a Yang-Mills theory where the one-dimensional electric field $E_{\theta}$ takes the place of $\Lambda$.    In \cite{Lano:1994gx,Henderson:1994vw,Branson:1996pe,Branson:1998bc,Gates:2001uu,Boveia:2002gf,Kilic} the authors argued that there was an analogous  field theory related to  the  dual of the Virasoro algebra that, like Yang-Mills, could be written in any dimension. Then one could lift the one-dimensional identity of $\mathcal{D}$ to a field theory in two-dimensions and higher.  In Section \ref{TWPolyakov} ,  we will show that by identifying  $\mathcal{D}$ with a  projective connection $\tilde \Gamma^{\alpha}_{\,\,\,\beta \gamma}$ that admits a projective Ricci tensor \({K}_{\a \b}\), the interaction term \ref{interact} can be written as   
\be
S_{\text{int}} = \int d^3x  \sqrt{{(-G)}} {K}_{\a \b}\, G^{\a \b}
 \ee  
where $G_{\a\b}$ will correspond to a specific three-dimensional metric  derived in Section \ref{metric} that contains the 2D metric \(g_{ab}\). This identification adds dimensional ubiquity to the diffeomorphism field \(\mathcal{D}_{ab} \).  Furthermore,  in Section \ref{TWdynamics} we postulate that  the    dynamical theory for $\mathcal{D}$ may be written as a projective version of the Gauss-Bonnet action, 
 \be
 S_{\text{{D}}} = \int d^3x  \sqrt{{(-G)}} \,\,{(K}^{\a}_{\,\,\,\b \g \rho}\,\,{K}_{\a}^{\,\,\,\b \g \rho}- \frac{1}{4}{K}_{\a \b}{K}^{\a \b}+ K^2),
 \ee
using the projective curvature tensor.  The three-dimensional metric $G_{\a \b}$ arises from  the chiral Dirac $\g^3$ matrix, which is used to define the  third dimension.  
 The construction is designed to be dimensionally independent and furthermore, the projective Gauss-Bonnet  action can be used in 2 to 4 space-time dimensions without introducing  higher time derivatives on the underlying metric.  In Section \ref{TWdynamics}, we will also derive  the field equations and energy-momentum tensor for any dimension. This work promotes the diffeomorphism field \(\mathcal{D}_{ab} \) from  a remnant of the dual of the Virasoro algebra to a rank two tensor in any dimension.  Its purely geometric origin  may have merits in describing gravitational and cosmological phenomena, especially as a candidate for dark energy and dark matter.  
  
\section{Review of Projective  Curvature }  
\subsection{Projective Structure and the Ricci Tensor}
Before discussing the results of this paper,  we review the relationship between projective structure and the Ricci tensor. The underlying idea is the equivalence between the family of geodesics on a manifold,  and the differential operators that give rise to the same geodesics.   In the study of \emph{sprays} (for a review see \cite{Crampin}), which may be viewed as the space of geodesics on a manifold and their derivatives, geodesic equations, Sturm-Liouville and Laplace equations are all placed on similar footing.  We are interested in the structure of geodesic equations under reparameterization.  For example in Eq.(\ref{2cocycle2}) one could ask how  the coadjoint element  transforms under reparameterization  so that the two-cocycle remains invariant. Sturm-Liouville theory investigates the same question about how the Liouville operator transforms under reparameterization of its parameter. These operations have a leading quadratic differential operator.      Indeed, Weyl \cite{Weyl} asked the same question regarding geodesics more than ninety years ago.  The idea of projective connections originated with Weyl, Thomas, Whitehead and Cartan \cite{Weyl,Whitehead,Thomas}.    It is born from the idea that  a manifold  $M$ can be characterized by its geodesics and geodetics (we will discuss the distinction presently).  Consider the geodesic equation,
\be 
\xi^a \nabla_a \xi^b = f \, \xi^b, \label{geodesic}
 \ee where $f$ is a function.  Here, the vector field $\xi$ has an affine parameter $\tau$, so that $\xi^a \nabla_a \tau =1.$   This geodesic equation has an inherent symmetry.  For one thing, geodesics and geodetics ($f=0$)  can be related by a suitable rescaling of $ g \,\xi^a =\, \zeta^a$. Indeed for any differentiable function $g$, $\zeta^a$ is also geodesic. In particular, when $ g$ satisfies $\frac{d }{d\tau}(\log{g(\tau))}= -f(\tau) $, $\zeta^a$ is geodetic.  We will refer to both geodetics and geodesics as geodesics from now on. 
Writing $\zeta^a = \frac{d}{d \tau} x^a$,  the geodesic equation in terms of the  affine connection $ \G^{a}_{\, b c}$ is 
\be
\zeta^b \nabla_b \zeta^a =\frac{d^2 x^a}{d\tau^2} + \G^a_{\, \, b c} \frac{d x^b}{d\tau} \frac{d x^c}{d\tau}=0. \label{geodesic2}
\ee
This demonstrates the symmetry of the affine parameter, $\tau\rightarrow c + b \,\tau$.
   The study of sprays through projective connections \cite{Crampin} addresses this symmetry.  

From the relationship between the affine connection  and the geodesics, one may ask, ``if given two  connections, when do they have the same family of geodesics?'' Weyl \cite{Weyl} had shown that if one considers two connections, say  $\hat \nabla$ and $\tilde \nabla$, then   they are projectively equivalent, i.e. admit the same family of geodesics,   if and only if there exists a one form $\omega$, such that 
\be
 {\hat \nabla}_a Y^b =  {\tilde \nabla}_a Y^b +\o_a Y^b + {\d_a}^b \o_m Y^m. \label{project} 
\ee    
These two connections are said to belong to the same equivalence class $[ \,{\hat \nabla}  \,] $ which is called a unique projective structure on $M  $.    Weyl was able to show the existence of a projective curvature, which is an invariant for each projective structure. 
Consider the Ricci tensor of two projectively related connections (see ``Notes on Projective Differential Geometry'' in \cite{2008SaOS}).  Explicitly, 
\bea 
&&({\tilde \nabla}_a {\tilde \nabla}_b  -  {\tilde \nabla}_b {\tilde \nabla}_a )Y^b = {\tilde R}_{a b}Y^b= {\tilde S}_{a b}Y^b -{\tilde A}_{a b} Y^b\\ &&\;\;\;\;\;\;\;\;\;\;\;\;\;\;\;\;\;\;\;\;\;\;\;\;\;\;\;\;\;\;\;\;\;\text{    and} \nonumber\\
&&({\hat \nabla}_a {\hat \nabla}_b  -  {\hat \nabla}_b {\hat \nabla}_a )Y^b ={\hat R}_{a b}Y^b = {\hat S}_{a b}Y^b -{\hat A}_{a b} Y^b,\nonumber
\eea
where $S_{a b}$ is symmetric and $A_{a b}$ is anti-symmetric.  Then the two-forms are related by,
\be  
{\hat  A}_{a b} = {\tilde A}_{a b} + {\tilde \nabla}_a \, \o_b -{\tilde \nabla}_b \, \o_a 
\ee
so  $A_{a b}$ has changed by an exact two-form.  The Bianchi identity shows that $dA=0$, where $d$ is the exterior derivative, therefore $A_{a b}$ is  closed.   This  implies a gauge symmetry when  $ A_{a b} \rightarrow A_{a b} + \partial_{[a}\L_{b]}$.  When    \(A_{ab}\) is exact,  a Levi-Civita connection is a member of the equivalence class. Since the Ricci tensor is  related to the Laplacian up to terms associated with the affine connection, the question asked originally regarding the two-cocycle is also being addressed here.  In this note, we will take advantage of the work on projective connections to elevate the one-dimensional coadjoint element to a field in two dimensions  and introduce  an action for the dynamics of the diffeomorphism field. 

\subsection{Projective Structure via Thomas and Whitehead}  
  
 There are many formulations of projective structures. We will follow the work of Thomas \cite{Thomas} and Whitehead \cite{Whitehead}.  For a review of projective connections, see \cite{Tabachnikov,OvsienkoValentin2005Pdgo,Crampin,2008SaOS}.    We may examine projective structures by treating the  diffeomorphism field, $\mathcal{D}$, as a  component of the projective connection $\tilde \Gamma^{\alpha}_{\,\,\,\beta \gamma}$ which is a \emph{connection on the space of affine connections}, $\nabla$. This may be formulated concretely by studying the family of affine connections on a manifold $\mathcal M$ with dimension $m$ through another connection, ``a connection of connections'', on a manifold $\mathcal N$ with dimension $n=m+1.$     The procedure below follows projective tractor calculus.
 
   Now $\mathcal N$ is to be regarded as the   projective space $ \mathbb{P}^m$ which is the quotient space of $ \mathbb{R}^{m+1}-{0}$ under multiplication by positive non-zero real numbers.    Let $\Upsilon^\a$ correspond to a vector field on $\mathcal{N}$ which  generates this action.     A function $f$ on $ \mathbb{P}^m $ then enjoys the symmetry generated by $\Upsilon$ so its Lie derivative vanishes, i.e.   $\mathcal{L}_\Upsilon f=\Upsilon^\a \partial_\a f=0$. In a Cartesian  frame, this is $x^\a \partial_\a = \Upsilon$.  In the projective connection below, we will realize \be \Upsilon^\a \partial_\a f= \l \,\partial_\l f , \label{lambda}\ee where $\l$ is the radial or ``volume'' parameter.  Furthermore we require that under parity,  ${\bf P}f(x)=f(-x).$   The parity restriction keeps us from making a change in affine parameters, as stated above, in such a way as to change an attractive force into a repulsive force. Functions such as this define the set $ {\mathcal F}_{\Upsilon}$.  Vector fields on  $ \mathbb{P}^m$ can be realized through their equivalence class defined by; 
\be \mathcal{L}_\Upsilon X \propto\ \Upsilon \ee  and ${\bf P} X = X.$  So $X \equiv Y$ if $Y-X \propto \Upsilon.$ We will use a construction that picks an element from each equivalence class as a representative of the equivalence class by its projection into a preferred one-form $\o$ on $ \mathbb{R}^{m+1}-{0}$ that enjoys $\bf{P}\, \o = \o$,  and is related to $\Upsilon$ by the conditions that  $\o_\a \Upsilon^\a=1  $ and $\mathcal{L}_\Upsilon \o_\rho =0$.  Explicitly, let $\tilde{X} \equiv X_{\text E}$  epitomize its equivalence class. Then $\tilde{X}^\b = X_{\text E}^\b - X_{\text E}^\rho \,\,\o_\rho \Upsilon^\b\ ,$ is equivalent to $X_{\text E}. $   Then the Lie algebra of equivalence classes can be defined through $[ X_{\text E},Y_{\text E}] = \mathcal{L}_{X_{\text E}}Y_{\text E} = Z_{\text E}.$  For functions in $ {\mathcal F}_{\Upsilon}$, say $g$, we have $\mathcal{L}_{Z_{\text E}}\,  g \in {\mathcal F}_{\Upsilon}. $   From this we can define a covariant derivative operator on the space of equivalence classes of vectors. 

The manifold $\mathcal N$ is  an example of a \emph{Thomas cone}, equipped with  an operator  \(  D(\tilde \Gamma^{\alpha}_{\,\,\,\beta \gamma}) \) is called the \emph{Thomas operator } \cite{BaileyT.N.1994TSBf}.
The construction allows us to have a notion of a gauge theory over the space of affine connections while preserving the idea of tensor representations. Indeed, Thomas \cite{Thomas} was able to relate a vector  bundle equipped with a connection to the irreducible representations of $SL(n+1,\mathbb{R})$ so that physics related to tensors even on the underlying manifold is maintained.    These are  the projective tractor bundles (for a review see Eastwood in \cite{2008SaOS} and \cite{curry_gover_2018}) which embed the projective transformations into its construction ab initio.
The ability to still discuss tensorality on the space  \( {\mathcal M} \) is what makes  an association with gravitation feasible.    
\subsection{Explicit Thomas-Whitehead Projective Connection}

As stated above, the Thomas-Whitehead theory may be thought of as ``gauging" the projective geometry by providing a connection as an \(\mathbb{R}_{+}\) fibration for projective transformations over the manifold \(\mathcal{M}\). To construct the covariant derivative operator explicitly,  we will need a parameter corresponding to flows  as defined by the principle vector $\Upsilon$.    This is achieved through the volume form,  $vol(\l) =  f(\l) \epsilon_{a_1 \cdots a_n}dx^{a_1}\cdots dx^{a_n}.$   Then $\l$ becomes the new direction parameter for $\mathcal{N}$ restricted to \(\mathbb{R}_{+}\).      This will be the definition used to describe $\l$ as stated in Eq.[\ref{lambda}]. Following \cite{Crampin,Roberts},   we construct the coefficients of the projective  connection $ {\tilde \nabla}_\a $ by requiring that

\be {\tilde \nabla}_\a \Upsilon^\b = \d_\a^\b . \label{primary}\ee
The new $m+1$ coordinates are $x^\a = (\l, x^1, x^2, \cdots x^m)$, where $\alpha= 0, \cdots, m.$  
Here the Greek letters denote the coordinates on ${\mathcal N}$ while the Latin correspond to the coordinates on $\mathcal M$  and are labeled $a,b = 1, \cdots, m$. The connection coefficients for  $\tilde \nabla$ on  $\mathcal N$ will be denoted by ${\tilde \G}^{\b}_{\,\,\,\rho \,\a}$ and the connection coefficients on $\mathcal M$ by  ${\G}^{b}_{\,\,\,c \,a}.$ Then in the frame  where \(\Upsilon^\a = (\l,0,0,\cdots,0)\) and \(\o_\b= (\frac{1}{\l},0, \cdots, 0)\)\cite{Roberts}, the projective connection coefficients are: 
\bea
{\tilde \G}^{0}_{\,\,\,a0}&=&{\tilde \G}^{0}_{\,\,\,0 a} = 0,\;\;\;\; {\tilde \G}^{0}_{\,\,\,\, a b} =  \Upsilon^0 {\mathcal{D}}_{ a b},\;\;\;\;{\tilde \G}^{\a}_{\,\,\,\,0 0} = 0,\cr &&\;\;\;\;{\tilde \G}^{a}_{\,\,\,\,0 b}={\tilde \G}^{a}_{\,\,\,\,b 0} = \o_0\,\d^a_b,  \; \text{and} \;\;\;\;{\tilde \G}^{a}_{\,\,\,\,b c} ={ \G}^{a}_{\,\,\,\,b c} ,
\eea 
   This choice guarantees that 
    \be {\bf P}\, \o_\b = \o_\b, \;\   \o_\a \Upsilon^\a=1   \;\text{   and   }\;\ \mathcal{L}_\Upsilon \o_\rho =0. \label{conditions2}\ee  
One then computes the projective curvature in the usual way, i.e 
\be 
[{\tilde \nabla}_\a,{\tilde \nabla}_\b] \kappa^\g = \,{K}^{\g}_{\,\,\,\a \b  \rho} \kappa^\rho. \label{projectivecurvature}
\ee
In terms of the connection coefficients we find,
\begin{align}\label{e:PCurv}
        K^\mu{}_{\nu\alpha\b} \equiv \tilde{\G}^\mu{}_{\nu[\b,\a]} + \tilde{\G}^\rho{}_{\n[\b}\tilde{\G}^\mu{}_{\a]\rho}~,
\end{align}
where the only non-vanishing components of $K^\mu{}_{\nu\alpha\b}$ are
\begin{align}
        K^a{}_{bcd} = R^a{}_{bcd} + \delta_{[c}{}^a \mathcal{D}_{d]b}~,&~~~ \label{Schouten}\\
        K^\l{}_{cab} = \lambda \partial_{[a} \mathcal{D}_{b]c} + \lambda \Gamma^d{}_{c[b}\mathcal{D}_{a]d}~= \l K_{cab} .&~~~ \label{Cotton}
\end{align}
Here one recognizes a familiarity with conformal structures on the manifold \( \mathcal M \). Recall that the Riemann curvature tensor and the Weyl tensor are related by
\begin{align}
        R^a{}_{bcd} = W^a{}_{bcd} + \delta_{[c}{}^a \mathcal{P}_{d]b}~,&~~~ \label{Weyl}\\
        C_{cab} =  \partial_{[a} \mathcal{P}_{b]c} +  \Gamma^d{}_{c[b}\mathcal{P}_{a]d}~,&~~~ \label{conformalCotton} 
\end{align}
where \( \mathcal P_{a b} \) is called the Schouten tensor and \( C_{cab} \) is called the Cotton-York tensor.  Under conformal transformations  of the metric, only the Schouten tensor will change.  Tractor calculus generalizes this to that in  
Eq.[\ref{Schouten}] we recognizes $\mathcal{D}_{ab}$  as precisely the  \emph{projective Schouten tensor} and Eq.[\ref{Cotton}] as  \( \l \) times the   \emph{projective Cotton-York tensor}, \(K_{bmn}= \nabla_m{\mathcal{D}}_{bn}-\nabla_n {\mathcal{D}}_{bm} \) \cite{CapA.2014Emip,GoverA.Rod2012DEgE,Gover:2014vxa}. 
 From the projective curvature tensor, we calculate the  components of the projective Ricci tensor $K_{\alpha\beta}\equiv K^\rho{}_{\alpha\beta \rho}$ to be,
\begin{align}
        K_{ab} & = R_{ab} - (m-1) \mathcal{D}_{ab}~~~,~~~K_{a\lambda} = \partial_\lambda\Gamma^b{}_{ab}=0.
\end{align}
In the above, $R^a{}_{bcd}$ may be taken as the  Riemann curvature tensor over the manifold $\mathcal{M}$, defined in terms of its Christoffel symbols $\Gamma^{a}{}_{bc}$.       From the Lagrangian point of view for general relativity, this need not be assumed when following  the Palatini formalism \cite{palatini}. There, the metric and the connection are treated as independent fields and  the  metric compatibility condition  follows from the  field equations of the  Einstein-Hilbert-Palatini action.  In  this work we assume the connection is compatible with the metric but the tractor calculus presented is sufficiently robust that it will handle the most general case.
 
Now under a projective transformation, the coordinates on \( \mathcal N \) transform as,
\begin{equation}
 p^\a=(\l, p^1, p^2, \cdots p^m) \rightarrow q^\a=(\l'+ J(q,p), q^1(p), q^2(p), \cdots q^m(p)),
 \end{equation}
 where $J(q,p)= |\frac{\partial q^i}{\partial p^j} |$ corresponds to the determinant of the Jacobian of the transformation of the coordinates on $\mathcal{M}$.
 One finds that projective invariance of the curvature, Eq.(\ref{projectivecurvature}),  induces the \emph{projective  gauge transformation} on  ${{\mathcal{D}}}_{a b}$   given by, 
\be
{{\mathcal{D}}'(q)}_{a b} = \frac{\partial p^c}{\partial q^a} \frac{\partial p^d}{\partial q^b}{{\mathcal{D}}(p)}_{c d} +  \frac{\partial p^l}{\partial q^c}( \frac{\partial^2 q^c}{\partial p^l \partial p^d} \frac{\partial^2 p^d}{\partial q^a \partial q^b }) + \frac{\partial q^m}{\partial p^{n}} \frac{\partial^3 p^n}{\partial q^m \partial q^a \partial q^b}.
 \ee
For an infinitesimal transformation in the direction of a vector field \(\xi^a\), the Lie derivative of \( \mathcal D_{ab} \) is given by 
\be
\mathcal{L}_\xi {{\mathcal{D}}(x)}_{c d} = \xi^a \partial_a {{\mathcal{D}}(x)}_{c d} + {{\mathcal{D}}(x)}_{a d} \partial_c \xi^a + {{\mathcal{D}}(x)}_{c a} \partial_d \xi^a + \partial_a \partial_c \partial_d \xi^a.
\ee
Thus there is a coordinate transformation worth of gauge symmetry corresponding to  \(m\) degrees of freedom. Since \({{\mathcal{D}}(x)}_{a b}\) is a symmetric tensor on $\mathcal M$, it will have a total of \(\frac{m(m+1)}{2}-m=\frac{m(m-1)}{2}\) degrees of freedom for \(m>1\) (\(m=1\) is the special case of the coadjoint elements of the Virasoro algebra where the projective invariant is defined by the two-cocycle).     We remark that the transformation properties in complex dimensions  can be found  in \cite{osgood1992,Molzon,Gunning1967}. 
Furthermore, Gunning \cite{Gunning1967} discusses the origin of the Schwarzian derivative in one complex dimension. 

\subsection{Schwarzian Derivative and Geodesics}

 From the point of view of the Thomas-Whitehead projective connection, the Schwarzian derivative enjoys ubiquity in any dimension through reparameterization of geodesics. To see this, consider a geodetic equation for a vector field $\xi^\a \equiv\ \frac{d x^\a}{d \tau}$ on $\mathcal{N}$:
\be
\xi^\alpha {\tilde \nabla}_\alpha \xi^\beta=0,
\ee
 where $\tilde \nabla$ is the Thomas operator. For the $a=1 \cdots m$ components,
\be
\frac{d^2 x^a}{d\tau^2} + \G^a_{b c} \frac{d x^b}{d\tau} \frac{d x^c}{d\tau}= f  \frac{d x^a}{d\tau}, \label{FirstTW}
\ee where $f =-2(\frac{d \log{(x^{0}(\tau))}}{d\tau}),$  and for the $\b=0$ component, 
\be
\frac{d^2 x^0}{d\tau^2} + x^0{\mathcal{D}}_{a b} \frac{d x^a}{d\tau} \frac{d x^b}{d\tau}=0. \label{SecondTW}
\ee 
Now changing the parameter,   $\tau \rightarrow \tau'(\tau)$, so that Eq.(\ref{FirstTW}) is also geodetic,   requires  
\be 
\frac{d^2 \tau'(\tau)}{d \tau^2} = -2 \frac{d \log{(x^{0}(\tau))}}{d\tau} \frac{d\tau'}{d\tau}. 
\ee Then performing this reparameterization on Eq.(\ref{SecondTW}) yields  \cite{Crampin}
\be 
 \frac{d}{d\tau}\left(\frac{\frac{d^2}{d \tau^2} \tau'(\tau)}{\frac{d\tau'}{d\tau}}\right)-\frac{1}{2}\left(\frac{\frac{d^2}{d \tau^2} \tau'(\tau)}{\frac{d\tau'}{d\tau}}\right)^2 = 2\, {\mathcal D}_{a b}\frac{d x^a}{d\tau} \frac{d x^b}{d\tau}. 
\ee The quantity
\be
\frac{d}{d\tau}\left(\frac{\frac{d^2}{d \tau^2} \tau'(\tau)}{\frac{d\tau'}{d\tau}}\right)-\frac{1}{2}\left(\frac{\frac{d^2}{d \tau^2} \tau'(\tau)}{\frac{d\tau'}{d\tau}}\right)^2= S(\tau',\tau) 
\ee
is precisely the Schwarzian derivative of $\tau'$ with respect to $\tau$  and is known to be invariant under M\"obius transformations, $SL(2,R)$, where \(\tau' \rightarrow\frac{a+ b \tau}{c+d \tau}\).   

\section{Thomas-Whitehead Projective Gravity}

 In what follows, the metric \(G_{\a \b}\) discussed above will be derived by using the chiral algebra of the Dirac matrices in even dimensions (see Section \ref{metric}) .  The resulting metric structure will then be used in any dimension. From there we can construct the spin-connection on \(\mathcal{N}\) and make contact with 2D Polyakov gravity through  fermions coupled to the projective geometry in Section \ref{TWPolyakov}.  Section \ref{TWchiral} will also discuss the possibility of chiral gravitational fields arising from distinct connections for the left-handed and right-handed fermions.  In Section \ref{TW2cocycle}, we will show that instead of using curvature invariants (which vanish in one-dimension), the projective Laplacian, \(G^{\a \b}\nabla_\a \nabla_\b\), over the one-dimensional circle  can be used to give rise to the 2-cocycle in Eq.(\ref{2cocycle2} ). Then in Section \ref{TWdynamics},  we will  describe dynamics for the field \(\mathcal{D}_{ab}\) in higher dimensions.  Since the projective curvature can be defined for any Riemannian manifold, we can define the dynamics through  a curvature squared type  action with \(G_{\a \b}\).  The projective Gauss-Bonnet action will be deemed the most suitable of the curvature squared actions, as our interest is mainly in 2,3 and 4 dimensions where the Riemannian Gauss-Bonnet action vanishes in 2 and 3 dimensions, and is a topological invariant in 4 dimensions. Thus higher derivative terms that might arise from curvature squared terms, such as the Kretschmann invariant density and Ricci squared, will not arise in these dimensions.  We explicitly derive the field equations for the diffeomorphism field \(\mathcal{D}_{ab}\), as well as its contribution to the energy-momentum tensor. 
We briefly examine the constraint dynamics and end with some additional discussion in Section \ref{conclusion}.

\subsection{The Metric  $G_{\a \b}$ and Spin Connection \({\tilde\o_\m}^{A B}\) }\label{metric} 
As stated above, the Thomas-Whitehead  connection  is defined on a manifold $\mathcal{N}$ that has one more dimension associated with its volume.   In general, there will not be a metric which is compatible with the connection.  Consider  even-dimensional manifold \(\mathcal{M}\). We define a metric on  $\mathcal{N}$ through the Clifford algebra of the Dirac matrices and the chiral Dirac matrix.   Let us assume that $\mathcal{M}$ admits spinors.  We can exploit the Dirac matrices to ``lift" a metric to  $\mathcal{N}$.    The Dirac matrices are related to a metric $g_{ab}$ on the manifold $\mathcal{M }$ via \be \{\g^a, \g^b\} = 2 g^{a b}.\ee 
  Here the coordinates on $\mathcal{M}$ are $a,b = 1,\cdots ,m.$ Then by using the volume form on \(\mathcal{M}\), we can define the chiral Dirac matrix, $\g_{m+1}$ (with new index down) that is related to the volume parameter $\l$  via 
\be
\gamma(\l)_{m+1} = \frac{f(\l)}{m!} i^{\frac{m-2}{2}} \epsilon_{a_1 \cdots a_m}\g^{a_1}\cdots \g^{a_m}.
\ee
   The intimacy of chirality and the volume form allows us to define a   metric on the manifold, $\mathcal{N}$, through
\be \{\g_\a, \g_\b\} = 2 G_{\a \b},\ee  
 where now  $\a,\b = 0,\cdots ,m.$  Then in any dimension, we may write the  metric \( G_{\a \b}\) as \begin{equation}
 G_{\a \b} =\begin{pmatrix}g_{a b} & 0 \\
0 & f(\lambda)^2 \\
\end{pmatrix}.
\end{equation}We will use $G_{\a \b}$ to contract with the  projective curvature tensor in the interaction Lagrangian and the dynamical action for the diffeomorphism field.
The volume of $G_{\a\b}$ is given by
 \be\sqrt{-\det{(G_{\a\b}})}= \sqrt{-\det{(g_{a b}})} f(\l). \ee  This metric coincides with metric projective tractors at a certain scale \cite{Gover:2014vxa,Branson:2004ap,Eastwood,GoverA.Rod2012DEgE,CapA.2014Emip,curry_gover_2018}.  To guarantee finite volume  when $\lambda$ is integrated from $0$ to $\infty$ in the action functionals, we choose $f(\l)$ so that    
\be
f(\l) = \exp(-2 \frac{\l}{\l_0}). \label{flambda}
\ee 
   Since  $G_{\a \b}$ admits frame fields through 
\be
G_{\m \n} = e_{\m}^{\,\,A} e_{\n}^{\,\,B} \eta_{A B}\,\,\,\, \text{and   } \eta_{A B} = g_{\m \n}E^{\m \,\,}_{\,\,A}  E^{\n \,\,}_{\,\,B},
\ee   
the projective spin connection may be written as
\be
 {\tilde\o_\m}^{A B} =e_{\n}^{ \,A} ({\tilde \nabla}_{\a}E^{\n B}).  
\label{spinconnection}  \ee 
The projective connection then acts on the gamma matrices as 
\be
\tilde{\nabla}_\mu \g^\nu = \partial_\mu  \g^\nu + [\Omega_\mu, \g^\nu] + \tilde{\G}^\nu_{\mu \sigma } \g^\sigma, 
\ee
and  
\be 
\Omega_\mu = \frac{1}{8} \o_\mu^{A,B} [\g_A, \g_B]
\ee
is the connection on fermions.  As we will remark later, in even dimensions, the fermion representation does not change in going from $m$ to $m+1 $ dimensions; there may be distinct projective connections for left and right handed chiral fermions.  This  introduces a natural notion of chiral symmetry in the gravitational sector.
\subsection{The 2-Cocycle and the TW Connection}\label{TW2cocycle}
As discussed in Section \ref{intro} , Eq.(\ref{correspondence}), Kirillov   was able to show a relationship between the Sturm-Liouville operator and the coadjoint elements of the Virasoro algebra.    By considering the Thomas-Whitehead over a one-dimensional manifold \( \mathcal{M} \) and using the metric \(G_{\a \b}\) we will present another way to see the correspondence. Here the conditions from Eq.(\ref{conditions2}) still hold and the Laplacian is used to construct the projective invariant. We consider a projective 2-cocycle on \(\mathcal{N} \) for a circle (this can also be defined on a line) given by 
\begin{equation}
<\xi,\eta>_{(\zeta)} = q \int_{C(\zeta)} \xi^\a ({\tilde \nabla}_\a G^{\rho \n} {\tilde \nabla}_\rho {\tilde \nabla}_\n \eta^\b\ G_{\b \m} )\zeta^\m d\s -(\xi \leftrightarrow \eta),
\end{equation}  where $\s$ parameterizes the path.  Here  the coordinates on  \(\mathcal{N} \) are $x^\a=(\l, x)$ and we choose the vector fields as $\xi^\alpha = (\xi_0,\xi_1)$ and $\eta^\a = (\eta_0,\eta_1)$. The vector $\zeta^\mu \equiv\frac{d x^\m}{d\s}$ defines the path $C$. In accord with Eq.(\ref{mapping}), the vector field  \(\eta\) is mapped into a quadratic differential, say \(\eta_{\a \m}\) via 
\be
(\tilde \nabla_\a G^{\rho \n} \tilde \nabla_\rho \tilde\nabla_\n \eta^\b G_{\b \m}) .
\ee For the sake of description, we write the one-dimensional metric as $g(x) $ and its Christoffel symbol as $\G$. Now consider a path given by a fixed value  $\l=\l_{0}$ along the vector,  $\zeta_{\l_0}^{\mu} = (0, \frac{dx}{d\s})$, and vector fields $\eta^\a = (0,\eta_1)$ and $\xi^\alpha = (0,\xi_1)$.   On such a path one finds that 
\begin{multline}
<\xi,\eta>_{(\zeta_{\l_0})} = q \int \xi_1 \left(2 \mathcal{D}_{11}+ g(x) \frac{1}{\l_0^2 f(\l_0)}+ 2 \G(x)^{2}+ \G'(x)\right)\eta'_1 dx \\
+ q \int \xi_1 \eta_1^{'''}dx -(\xi \leftrightarrow \eta).
\label{projective2cocycle}\end{multline}        
The term $2 \G(x)^{2}+ \G'(x) $ dictates the transformation law for  $\mathcal{D}_{11}$ and is algebraically equivalent to Eq.(\ref{variation_D}). Also, the metric is a tensor and has no inhomogeneous transformation law.  We take the metric to be the standard metric, \(g(x)=1\), so that Eq.(\ref{projective2cocycle}) becomes
\begin{equation}
<\xi,\eta>_{(\zeta_{\l_0})} = q \int \xi_1 \left(2 \mathcal{D}_{11}+  \frac{1}{\l_0^2 f(\l_0)}\right)\eta'_1 dx+ q \int \xi_1 \eta_1^{'''}dx -(\xi \leftrightarrow \eta).
\end{equation} 
Since $\l_0$ is fixed, we make the identification that  \(2\, q\,  \mathcal{D}_{11}+  \frac{q}{\l_0^2 f(\l_0)} =  B,   \) which recovers Eq.(\ref{2cocycle2}) for \(q=\frac{c}{2 \pi}\).        
 \subsection{Polyakov-Diffeomorphism Field Interaction Term}\label{TWPolyakov}
The Polyakov metric  will determine the $m=2$  dimensional metric on $\mathcal{M}$,  and we will extend it to a metric on $\mathcal{N}$ via the matrix $\g^3$ as mentioned above.      
 Explicitly, the Polyakov metric\cite{Polyakov:1981rd,DANIELSSON1989292} is  
\be  g_{a b} = \begin{pmatrix}0 & \frac{1}{2} \\
\frac{1}{2} & h_{\tau \tau}(\theta,\tau) \\
\end{pmatrix},\ee
where  $h_{\tau \tau}(\theta, \tau)= \frac{\partial_\theta\ f(\theta,\tau)}{\partial_\tau f(\theta,\tau)}$.   With this choice of metric,  the $\g^3$ extended metric is 
\be 
G_{\a \b} = \begin{pmatrix}0 & \frac{1}{2} & 0 \\
\frac{1}{2} &  h_{\tau \tau} & 0 \\
0 & 0 & f(\l)^2 \\
\end{pmatrix}.
\ee

The projective curvature coefficients $K_{\;\;\a \b \g}^{\rho}$ for the Polyakov metric on $\cal{M}$ are given below:   

\be
K^{3}_{\,\,\,a b c}= \begin{cases}
   \l(- \, \partial_\tau \mathcal{D}_{\theta \theta} +  \partial_\theta \mathcal{D}_{\theta \tau} + \mathcal{D}_{\theta \theta}\, \partial_\theta h_{\tau \tau})    ,& {a=1, b=1,c=2 } \\   -\l( \, -\partial_\tau \mathcal{D}_{\theta \tau} + \, \partial_\theta \mathcal{D}_{\tau \tau} -  2 \,\mathcal{D}_{\theta \tau}\, \partial_\theta h_{\tau \tau}\\\;\;\;\;\;\;\;\;\;\;\;\;\ + \mathcal{D}_{\theta \theta} (\partial_\tau h_{\tau \tau}+ 2(\partial_\theta h_{\tau \tau}\, h_{\tau \tau} )) ,& {a=2, b=1,c=2 } \\ 0,              & \text{otherwise}   
\end{cases}
\ee
\be
K^{2}_{\,\,\,a b c}= \begin{cases} 
-\mathcal{D}_{\theta \tau} -\partial^2_\theta h_{\tau \tau},      & {a=2, b=1,c=2} \\
  -\mathcal{D}_{\theta \theta},      & {a=1, b=1,c=2} \\0,              &
   \text{otherwise}   \end{cases}
\ee  

\be
K^{1}_{\,\,\,a b c}= \begin{cases} 
\mathcal{D}_{\theta \tau}+ \partial^2_\theta h_{\tau \tau},      & {a=1, b=1,c=2}\\\mathcal{D}_{\tau \tau}+ 2 h_{\tau \tau}\,  \partial^2_\theta h_{\tau \tau},      & {a=2, b=1,c=2} \\ 0,              & \text{otherwise}.  \end{cases}
\ee  

The projective-Ricci tensor then is computed via $K_{\;\;\a \b \rho}^{\rho} = K_{\a \b}= K_{\b \a }$  and has coefficients
\be
K_{\a \b }= \begin{cases} 
 \mathcal{D}_{\theta \theta}    & {\a=1, \b=1} \\
-\mathcal{D}_{\theta \tau} - \partial^2_\theta h_{\tau \tau},              & \a=1, \b=2 \\
-\mathcal{D}_{\tau \tau} -2 h_{\tau \tau} \partial^2_\theta h_{\tau \tau}, &  \a=2, \b=2 . \end{cases}
\ee  

Using the metric $G_{\a \b}$ on $\cal{N}$, we can find  the  interaction term between the projective connection and the induced Polyakov metric through 
\bea 
S_{\text{Diff Inter}} &=& \int\sqrt{-G} G^{\m \n} K_{\mu \nu}  d\l\; d\theta\; d\tau
\cr  &=& \int_{}^{}d\l d\theta d\tau \, \frac{f(\l)}{\sqrt 2}\, ( \mathcal{D}_{\theta \theta}\, h_{\tau \tau }-\mathcal{D}_{\theta \tau} -\partial^2_\theta h_{\tau \tau}), 
\cr  &=& \int_{}^{} d\theta d\tau \,\left(\int d\l\frac{f(\l)}{\sqrt 2}\right)\, ( \mathcal{D}_{\theta \theta}\, h_{\tau \tau }-\mathcal{D}_{\theta \tau} -\partial^2_\theta h_{\tau \tau})\eea
where the last term is  the  scalar curvature  and  is a total derivative.    Similarly, the middle term  integrates to a constant as it decouples from the interaction with the metric.  Therefore we have  the interaction term found in \cite{Rai:1989js}, 
\be
S_{\text{Diff Inter}} =q \int d\theta\, d\tau {\mathcal{D}}_{\theta \theta}\, h_{\tau \tau },
\ee
where $q$ is a constant. Thus, using the projective connection over two-dimensional metrics in the Polyakov gauge, we are able to recover the interaction term, Eq.(\ref{interact})    between the Polyakov metric and the coadjoint element.    

\subsection{Chiral Gravity through Projective Geometry}\label{TWchiral}
One  could have recovered this interaction term by computing  the chiral anomaly with a Dirac operator defined by the projective connection.   Since the dimension of the irreducible representation of the fermions does not change in going from $n$ to $n+1$ dimensions via the matrix $\g^{n+1}=\g^{\text chiral}$, one can arrive at this result by computing $\text{Tr}(\g^3)$ which written in invariant form  on the projective space is \be\g^3=\frac{\sqrt{2}}{\sqrt{\Upsilon^\a \Upsilon_\a}}\text{Tr}(\Upsilon^\a \g_\a).\ee 
We couple  left handed fermions to a Dirac equation by using the projective Dirac operator ${\tilde{ \slashed{\nabla}}}$ in the heat kernel expansion. There   the square of the Dirac operator yields  
\be
{\tilde{ \slashed{\nabla}}}{\tilde{ \slashed{\nabla}}}= \g^\a {\tilde \nabla}_{\a}(\g^\b) {\tilde \nabla}_{\b} + (G^{\a \b} -i \sigma^{\a \b}){\tilde \nabla}_{\a}{\tilde \nabla}_{\b}.
\ee 
The requisite term to computing the trace is 
\be 
\sigma^{\a \b}{\tilde \nabla}_{\a}{\tilde \nabla}_{\b} =\frac{1}{2}\sigma^{\a \b}[{\tilde \nabla}_{\a},{\tilde \nabla}_{\b}] =\frac{1}{2}\sigma^{\a \b}{\bf  K}_{\a \b}=\frac{1}{2}\sigma^{\a \b}\frac{1}{2}\sigma_{\d \l}{ K}_{\a \b}^{\,\,\,\,\,\,\,\,\d \l} \propto { K}  ,
\ee
where ${\bf K}_{\a \b} $ is the projective curvature two-form constructed from the projective spin connection Eq.(\ref{spinconnection}).  Since $\g^{m+1}$ has two eigenvalues that correspond to distinct orientation forms, we may write a chiral gravitational theory for fermions in even dimensions by adding  a Dirac action on the projective space.  Then   by introducing two distinct projective connections for the vector and axial vector, we can write a chiral action as  
\be 
S_{(\psi,\mathcal{D}^V,\mathcal{D}^A)} = \int d^{n+1}x  \sqrt{\det{(-G)}} i\left( {\bar \psi}\g^\mu \tilde{{\nabla}}^V_\mu \psi + {\bar \psi}\g^\mu \tilde{\nabla}^A_\mu \text{}(\Psi^\a \g_\a) \psi \right).
\ee The advantage of this approach is that we do not have to introduce separate metrics for the chiralities, since  \(\mathcal{D}^\text{vector}_{ab}\) and  \(\mathcal{D}^\text{axial}_{ab}\) are independent fields that have distinct isotropy groups.  When the two diffeomorphism fields are the same, chiral symmetry is broken. 
We are presently investigating how such a coupling affects chiral symmetry and chiral symmetry breaking through gravity.    
\subsection{Projective Curvature Squared and Diffeomorphism Field Lagrangian}\label{TWdynamics}
\subsubsection{Thomas Whitehead Projective Gravitational Action}
Now we turn our attention to constructing a Lagrangian for the diffeomorphism field in order to describe it as a dynamical field in its own right.   Motivated by the discussion in the previous section, a Lagrangian for the diffeomorphism field could be of the form

\be
S_{\text{trial action}} =  \int d^n x d\lambda \sqrt{-G}\,  K_{\;\;\a \b \g}^{\rho} K_{\;\;\m \n \s }^{\d} G^{\a \m}G^{\b \n}G^{\g \s}G_{\rho \d} . \label{diffaction}
\ee
However, an expansion of this action includes a Kretschmann scalar density, 
\begin{displaymath}
 \int d^n x  \sqrt{-g}\,  R_{\;\;b c d}^{a} R^{\;\;b c d }_{a}. 
\end{displaymath} Such a term will give rise to higher derivative terms in the metric.  Furthermore, the action is not explicitly projectively invariant.  Since the Gauss-Bonnet action is trivial in two and three dimensions and a topological invariant in four dimensions, we are motivated to write our action as 
  
\begin{eqnarray}
&&S_{\text {TWPG}} = \alpha_0 \int d^m x d\lambda \sqrt{-\det{(G_{\mu \nu})}}K_{\alpha  \beta} G^{\alpha \beta} + S_{\text PGB},\,\,\,\,\, \text{where} \label{PEH}\\
&&S_{\text {PGB}} = \beta_0 \int d^m x d\lambda \sqrt{-\det{(G_{\mu \nu})}}\left(K^2-4 K_{\mu \nu}K^{\mu\nu} + K_{\mu \nu \rho}^{\,\,\,\,\,\,\,\,\,\alpha}  K^{\mu \nu \rho}_{\,\,\,\,\,\,\,\,\,\alpha}  \right), \label{PGB}
\end{eqnarray}
where \(K_{a b} = R_{a b} -(m-1) \mathcal{D}_{a b}\) and  \(K = R -(m-1) g^{a b} \mathcal{D}_{a b}=R-(m-1)\mathcal{D}.\) We will call this the  Thomas-Whitehead projective gravitational action. This may be expanded as  the $m$-dimensional action
\begin{eqnarray} 
S_{\text {TWPG}}  &=&\alpha_0 \frac{\lambda_0}{2} \int_{}^{}\,d^m x   \sqrt{-g} g^{ab} R_{a b} \cr 
&-&(m-1)\alpha_0 \frac{\lambda_0}{2}\int_{}^{}\,d^m x  \sqrt{-g} g^{ab} {\mathcal{D}}_{a  b} \cr
&-&\frac{\beta_0\lambda_0}{2} \int d^m x \sqrt{-g}\left(4 R_{ab}{\mathcal{D}}^{ab} -2 (m-1) {\mathcal{D}}_{ab}{\mathcal{D}}^{ab} - \frac{\lambda_0^2}{54} K_{bmn}K^{bmn} \right)\cr
&+&2(m-1) \beta_0\lambda_0 \int d^m x \sqrt{-g} \:{\mathcal{D}}_{ab}\left(2R^{ab} - (m-1){\mathcal{D}}^{ab}\right)\cr
&-&( m-1) \frac{\beta_0}{2}\lambda_0 \int d^m x \sqrt{-g}{\:\mathcal{D}}\left(2 R - (m-1){\mathcal{D}}\right)+S_{\text GB},
\end{eqnarray}
where the $\l$ integration has been performed using Eq.(\ref{flambda}).   Here $\a_0 \equiv {\kappa^2}$ where $\kappa$ is the gravitational constant.  The quadratic Gauss-Bonnet action, 
\be S_{GB}= \frac{\beta_0}{2}\lambda_0\int d^n x  \sqrt{-g}\left(R^2-4 R_{a b}R^{a b} + R_{a b c}^{\,\,\,\,\,\,\,\,\,d}  R^{a b c}_{\,\,\,\,\,\,\,\,\,d}  \right),\ee
vanishes in two and three dimensions and is a topological term in four dimensions.   The dynamics for ${\mathcal{D}}_{ab} $ arise from the term $K_{bmn}K^{bmn}$ where $K_{bmn}= \nabla_m{\mathcal{D}}_{bn}-\nabla_n {\mathcal{D}}_{bm}$ is the projective York-Cotton tensor.  This satisfies the Bianchi identity 
\be K_{acm} + K_{cma} + K_{mac} = 0.\ee The field equations for ${\mathcal{D}}_{ab} $ that arise from this action are
\begin{equation}
\beta_0 (\frac{\lambda_0}{3})^3 \nabla_m( K^{cdm} + K^{dcm}) - \lambda_0 (m-1) (\alpha_0 + 2\beta_0 K) g^{cd} + 4 \beta_0 \lambda_0 (2m-3) K^{cd}=0.
\end{equation} This can be written in the form \(\mathcal{O}^{c d;a b} \mathcal{D}_{ab} = J^{cd},\) where \(\mathcal{O}^{c d;a b}\) is a second order differential operator and \( J^{cd}\) a metric dependent source, i.e.
\begin{multline}
\beta_0 (\frac{\lambda_0}{3})^3 \nabla_m( K^{cdm} + K^{dcm})  - 4 \beta_0 \lambda_0 (2m-3)(m-1)\mathcal{D}^{cd} + 2\lambda_0 \beta_0(m-1)^2 \mathcal{D} g^{cd} \\ =-4 \beta_0 \lambda_0 (2m-3)R^{cd}+ \lambda_0 (m-1) (\alpha_0 + 2\beta_0R) g^{cd}.
\end{multline}

In Thomas-Whitehead projective gravity, the field \(\mathcal{D}_{ab}\) is physical  and can back-react with the metric.  By varying the action with respect to the metric \(g_{ab}\), we  can write the Einstein equations as
\begin{equation}
\kappa^2 (R^{lm}-\frac{1}{2}R g^{lm})=\Theta^{l m}+\Theta_{GB}^{l m},
\end{equation}
where we have separated out the energy-momentum contribution from the Riemannian Gauss-Bonnet term as \(\Theta_{GB}^{l m}\).  The contribution due to the diffeomorphism field is given by  \(\Theta^{l m}\). This can be written in four parts corresponding to  the contribution arising from the projective Einstein-Hilbert action, which is the first summand in Eq.(\ref{PEH}),
and the three terms in Eq(\ref{PGB}) arising from the projective scalar squared term (PSS),    the projective curvature squared term (PCS) and the the projective Ricci squared term (PRS).  Using these labels we write
\begin{eqnarray}
&&\Theta^{m n} =\Theta^{lm}_{\text {PEH} }+\Theta^{m n}_{\text {PSS} } +\Theta^{m n}_{\text {PCS} }+ \Theta^{m n}_{\text{ PRS }} \\
&& {\text {where}} \cr
&& \Theta^{lm}_{\text {PEH} } =   +\frac{1}{2}\kappa^2 (n-1)({\mathcal{D}}^{lm}-\frac{1}{2}{\mathcal{D}} g^{lm}),  \\ \cr
&&\Theta^{lm}_{\text {PSS} }= 2\beta_0 \lambda_0 (1-m)(g^{lm} g^{cd}  - g^{lc} g^{md})\nabla_c \nabla_d\mathcal{D}    + \beta_0 \frac{\lambda_0}{2} (2(1-m) R \mathcal{D}\cr && + (1-m)^2 \mathcal{D}^2) g^{lm} -2 \beta_0 \lambda_0  (1-m)(R^{lm} \mathcal{D}+\mathcal{D}^{lm} R+(1-m)\mathcal{D}^{lm} \mathcal{D}), \cr &&\\
&& \Theta^{lm}_{\text {PCS} }= \beta_0 \lambda_0 \left( {\mathcal {L}} + 4(R^l_{\,\,c}{\mathcal{D}}^{mc} + R^m_{\,\,c}{\mathcal{D}}^{cl}) -2(n-1)({\mathcal{D}}^l_{\,\,c}{\mathcal{D}}^{mc} + {\mathcal{D}}^m_{\,\,c}{\mathcal{D}}^{cl}) \right) \cr
&& -\beta_0 \frac{\lambda_0^3}{54}\,\ K^l_{\,\,\,\,ca} K^{mca}  -\beta_0 \frac{\lambda_0^3}{27} \,K^{cdl} K_{cd}^{\,\,\,\,\,\,m} -\beta_0 \lambda_0\nabla_c X^{clm},\\ \cr
&&\Theta^{lm}_{\text {PRS} }= -2 \beta_0 \lambda_0 (1-m) g^{l m}\left(  2R_{ab} \mathcal{D}^{ab} +(1-m)\mathcal{D}_{ab} \mathcal{D}^{ab} \right)\cr &&\,\,\,\,\,\,\,\,\,\,\,\,-4(1-m)(R^{cl} \mathcal{D}_c^{\,\,m}+\mathcal{D}^{cl} R_c^{\,\,m}+(1-m)\mathcal{D}^{cl} \mathcal{D}_c^{\,\,m})\cr &&\,\,\,\,\,\,\,\,\,\,\, +2(1-m)\nabla_a\nabla_b (\mathcal{D}^{ab} g^{lm} +\mathcal{D}^{lm} g^{ab}-\mathcal{D}^{mb} g^{la} -\mathcal{D}^{al} g^{mb})\\
\text{Here}\cr
&& X^{cab}=J^{cab}-J^{abc}-J^{bac},\;\   \\
&&J^{cab} = g^{ca}\nabla_m\mathcal{D}^{mb} +g^{cb}\nabla_m\mathcal{D}^{ma} - 2 g^{cd} \nabla_d \mathcal{D}^{ef}\cr  &&\;  \,\,\,\,\,\,\,\,\,\,\,-\frac{\lambda_0^2}{54}(\mathcal{D}_m^{\,\,c} K^{abm} + \mathcal{D}_m^{\,\,c}K^{bam})\;\;\\ \text{and}\nonumber  \\ &&\mathcal{L} =  -\frac{\lambda_0^2}{54}(K_{bmn}K_{acd})g^{ab}g^{mc}g^{nd}-2(n-1)\mathcal{D}^{ab}\mathcal{D}_{ab} +4 R^{ab}\mathcal{D}_{ab}. 
\end{eqnarray} 
For the sake of completeness, in dimensions higher than four, the Gauss-Bonnet action contributes to the total energy-momentum tensor through  divergent-free  Lanczos tensor \cite{Lanczos:1938sf}, 
\begin{multline}
\frac{1}{\b_0 \l_0}\Theta_{GB}^{l m}= \\\frac{1}{2}g^{lm}(R^2 - 4 R_{a b} R^{a b} + R_{abcd} R^{abcd}) -2 R R^{lm} - 4 R^{ablm}R_{ab} - 2 R^{lbcd}R^{m}_{\:\:\:b c d} + 4 R^{a l} R^{m}_{\;\;\;a}. 
\end{multline}
\section{2D TWPG}
Here we briefly discuss the relevant degrees of freedom in the  ``free'' Lagrangian in 2D. For this we do not include the projective Einstein-Hilbert action and explicitly use a metric where   $\G^a_{\;\;bc}=0$. From here we can examine salient features of  the phase space structure of the action.  An  in-depth analysis will be carried out through Dirac brackets in a future work.     For this simplified case,  we have the 2D projective action 
\bea
\begin{split}
S_{\text{ Diff free}} = - \a_1 \int d\theta  d\tau  \left( ( {\mathcal{D}}_{\theta \theta} +  {\mathcal{D}}_{\tau \tau})^2- 4{\mathcal{D}}_{\theta \tau}^2 \right)\\ -\a_2 \int d\theta  d\tau \left( (\partial_\tau \mathcal{D}_{\theta \tau} -\partial_\theta \mathcal{D}_{\tau \tau})^2 -(\partial_\tau \mathcal{D}_{\theta \theta} -\partial_\theta \mathcal{D}_{\theta \tau})^2\right) ,
\end{split}
\eea
where 
\be \a_1= \int \frac{1}{\sqrt{2}} f(\l) \;\;\text{and}\;\;  \a_2 = \int {\sqrt{2}} \l^2 f(\l)^3.  \ee
The  conjugate momenta are given by
\bea
&&\Pi_{\theta \tau}= -2\a_2 ( \partial_{\tau} \mathcal{{D}}_{\theta  \theta}- \partial_{\theta} \mathcal{{D}}_{\theta  \tau} )\cr
 &&\Pi_{\theta \theta}= +2 \a_2( \partial_{\tau} \mathcal{D}_{\theta  \tau}- \partial_{\theta} \mathcal{D}_{\tau  \tau} )\cr
 &&\Pi_{\tau \tau}=0,
 \eea
where the  vanishing of $\Pi_{\tau \tau}$ signifies a primary  constraint, $\Phi_1 = \Pi_{\tau \tau}$.  The equations of motion for $\mathcal{D}_{\tau \tau} $ then give a secondary constraint $\Phi_2$, where   
\be
\Phi_2 =- \sqrt{2} \a_1 ({\mathcal{D}}_{\theta \theta}+{\mathcal{D}}_{\tau \tau})\ + \partial_\theta \Pi_{\theta \tau.}\ee
In terms of the conjugate momenta, the  field equations for the three fields $\mathcal{D}_{\theta \theta}, \mathcal{D}_{\theta \tau},$ and $\mathcal{D}_{\tau \tau}$ respectively are
\bea
&& \sqrt{2} \a_1 ({\mathcal{D}}_{\theta \theta} +{\mathcal{D}}_{\tau \tau})- \partial_\theta \Pi_{\theta \tau}=0 \cr
&& \sqrt{2}\a_1  {\mathcal{D}}_{\theta \tau}  + \,\partial_\theta \Pi_{\theta \theta} =\partial_\tau \Pi_{\theta \tau}\cr &&\sqrt{2}\a_1( {\mathcal{D}}_{\theta \theta} +{\mathcal{D}}_{\tau \tau})   + \,  \partial_\tau \Pi_{\theta \theta}=0. 
\eea

Here one may view the $\a_1$ terms as mass terms, demonstrating that  masses are derived from the   projective geometry.  The structure of the constraint algebra for Dirac quantization will depend on whether ${\mathcal{D}}_{\tau\tau}$ vanishes, which will lead to first class constraints via Dirac brackets.   Only one physical degree of freedom will remain. We expect that known techniques for handling partially massless gauge theories \cite{Gover:2014vxa} will also be of value here. This is under investigation.

\section{Conclusion}\label{conclusion}

We have shown that projective geometry may be used to give dynamics and a field theoretic interpretation to the coadjoint elements which arise in the study of 2D gravity.  The fixed coadjoint orbits are interpreted  as  external energy-momentum tensors in the absence of dynamics.   For orbits such as $\text{Diff} S^1/SL_n$ with $\mathcal{D}_{ab} \propto -n^2 $, the coadjoint element acts as a negative cosmological constant in $2D$.       Furthermore, the projective curvature renders  the diffeomorphism field a dynamical field theory, putting the orbit construction in the Virasoro sector on the same footing as the affine Kac-Moody orbits, where the coadjoint elements can be regarded as background Yang-Mills fields.   The construction uses the property  of  coadjoint elements  as projective structures on $S^1$  which allows us to  ``lift'' the  projective connection to any dimension.  In particular we are able to identify the diffeomorphism field \( \mathcal D_{ab} \) with the projective Schouten tensor found in Tractor calculus. By introducing a metric structure associated with the chiral Dirac algebra, we can compute curvature squared invariants and  provide dynamics to the projective connection.  Using the Gauss-Bonnet action, we can compute an energy-momentum tensor for the diffeomorphism field and avoid higher derivatives on the metric for dimensions less than five.  This energy-momentum tensor demonstrates that the diffeomorphism field will couple to matter gravitationally.  A fully back-reacted solution will modify geodesics and will appear as a source of dark energy.  Whether this is related to phenomenologically observed dark energy and dark matter will depend on the \(\b_0\) coupling constant. This is  presently under investigation.  Indeed, one could fix the affine connection due to general relativity and use the projective connection to study perturbations which give distinct geodesics.  As suggested in \cite{Branson:1996pe}, gravitation may be equipped with not only a metric but another field so that the pair $(g_{ab}, {\mathcal{D}}_{\a \b})$  describes the dynamics.  We now recognize that this is the metric and the projective Schouten tensor that can be associated with specific tractor bundles. Dynamical projective connections may open a new avenue to the understanding of cosmology, gravitational radiation, and the gravitational dynamics of matter. However, much work is left to be done to pin down the phenomenology of this approach.

\section*{Acknowledgments}
The authors thank  Chris Doran, Kenneth Heitritter, Delalcan Kilic, Calvin Mera,  and the members of the Nuclear and High Energy Theory group at the University of Iowa for discussion.  We especially thank Kory Stiffler for carefully reviewing the manuscript. V.R. especially thanks the late Prof. Thomas Branson for discussion in  projective geometry that initiated this work. The authors would also like to thank the referee for directing us to the literature in tractor calculus.

\bibliographystyle{JHEP}
\bibliography{PaperProjective2D}
\end{document}